\documentclass[prl,superscriptaddress,twocolumn,showpacs]{revtex4}
\usepackage{amsmath,graphicx}

\begin{document}
\title{Very High Frequency Spectroscopy and Tuning of a Single-Cooper-Pair-Transistor with an on-chip Generator}
\author{P.-M. Billangeon}
\affiliation{Laboratoire de Physique des Solides, Univ. Paris-Sud, CNRS, UMR 8502, F-91405 Orsay Cedex, France.}
\author{F. Pierre}
\affiliation{Laboratoire de Photonique et Nanostructures, CNRS, UPR 20, F-91460 Marcoussis, France.}
\author{H. Bouchiat}
\affiliation{Laboratoire de Physique des Solides, Univ. Paris-Sud, CNRS, UMR 8502, F-91405 Orsay Cedex, France.}
\author{R. Deblock}
\affiliation{Laboratoire de Physique des Solides, Univ. Paris-Sud, CNRS, UMR 8502, F-91405 Orsay Cedex, France.}
\pacs{73.23.-b, 73.23.Hk, 74.50.+r, 78.70.Gq}
\begin{abstract}
A single Cooper pair transistor (SCPT) is coupled capacitively to a voltage biased Josephson junction, used as a high frequency generator. Thanks to the high energy of photons generated by the Josephson junction, transitions between energy levels, not limited to the first two levels, were induced and the effect of this irradiation on the DC Josephson current of the SCPT was measured. The phase and gate bias dependence of energy levels of the SCPT at high energy is probed. Because the energies of photons can be higher than the superconducting gap we can induce not only transfer of Cooper pairs but also transfer of quasi-particles through the island of the SCPT, thus controlling the poisoning of the SCPT. This can both decrease and increase the average Josephson energy of the SCPT~: its supercurrent is then controlled by high-frequency irradiation.
\end{abstract}
\maketitle

Circuits based on small Josephson junctions can behave like macroscopic quantum systems \cite{Fulton89,Martinis87,Joyez94,JoyezPhD}. This is in particular the case for the single Cooper pair transistor (SCPT), a superconducting metallic island connected \textit{via} tunnel junctions to two superconducting reservoirs. The electrostatic state of the island is controlled by a nearby gate. Due to the interplay of charging and Josephson effect this system can be considered, at low voltage, as a Josephson junction which Josephson energy, and thus critical current, can be tuned by a gate voltage. Due to its fundamental interest as an electrometer and as a building block for quantum computing the SCPT has been extensively studied over the last 15 years. To resolve its gate voltage and superconducting phase dependent energy levels essentially two techniques were used. The first is spectroscopy by using microwave irradiation on the gate, often limited to transitions between the first two energy levels. The second technique is transport spectroscopy~: current is measured at different bias voltages $V_B$ and gate voltages $V_G$ to probes energy levels within $e V_B$ of the ground state. The current results from all the tunneling processes allowed at energy $e V_B$, leading to an interesting but complicated characteristic $I(V_B,V_G)$.

In this article we investigate a SCPT irradiated by high frequency (HF) photons generated by a Josephson junction. This allows us to work at frequencies much higher than usual spectroscopy experiments. We study how the average Josephson coupling of the SCPT is affected by HF irradiation. Because this technique permits to probe transitions involving only a single Cooper pair (CP) or a single quasiparticle (QP) it gives very direct information on the high energy spectrum of the SCPT, in particular on the tunneling of one QP to or from the island, i.e. the poisoning of the SCPT, an important issue in the context of quantum computing with states involving only CPs.

\begin{figure}
	\begin{center}
		\includegraphics[width=7.5cm]{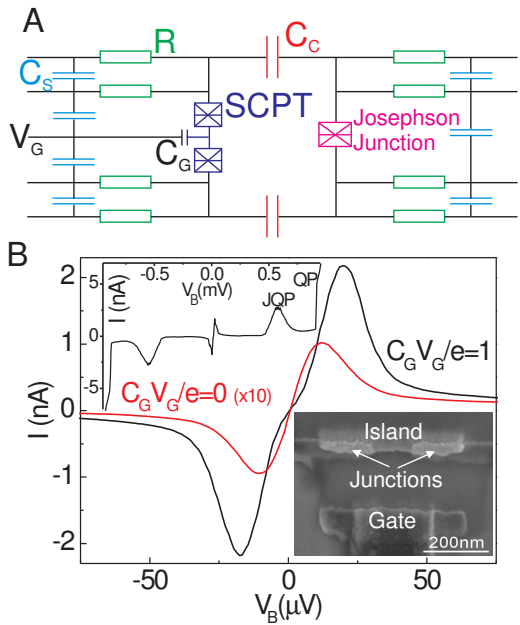}
	\end{center}
	\caption{A : Sketch of the SCPT coupled to a Josephson junction and the on-chip circuit ($R=750$ $\Omega$, $C_C \approx 750$ fF). B : Josephson branch of the SCPT at $C_G V_G/e=1$ and $C_G V_G/e=0$ (multiplied by 10). Upper inset : $I(V)$ characteristic of the SCPT at high bias showing the JQP peak and the QP tunneling. Lower inset : SEM picture of the SCPT.}
	\label{fig:figure1}
\end{figure}

The device probed is a SCPT (normal state resistance 48.5 k$\Omega$) coupled capacitively to a small Josephson junction (estimated capacitance 1 fF, normal state resistance 25 k$\Omega$). Both structures are made of aluminum (superconducting gap $\Delta$=210 $\mu$eV) and embedded in an on-chip environment consisting of resistances (8 Pt wires, $R$=750 $\Omega$, length=40 $\mu$m, width=750 nm, thickness=15 nm) and capacitances (estimated value $C_C \approx 750$ fF, size~: 23$\times$25 $\mu$m$^2$, insulator~: 65 nm of Al$_2$O$_3$) designed to provide a good high-frequency coupling between the two devices (Fig. \ref{fig:figure1}A) \cite{deblock03,billangeon06}. The sample is measured in a dilution refrigerator of base temperature 90 mK. 

We first present transport measurements on the SCPT. Fig. \ref{fig:figure1} shows the $I(V_B)$ characteristic for low voltage of the SCPT at two values of the gate voltage. The SCPT exhibits a dissipative Josephson branch which extends to finite voltages \cite{NoteJunction} and with a finite slope $dI/dV_B$ at low bias voltage. This is commonly seen on SCPTs \cite{Kycia01,Lu02,Lotkhov03} (and Josephson junction \cite{Averin01,Kuzmin91,Ingold92,Ingold99}) when they are embedded in a dissipative environment. The effect of the environment is  more important if the Josephson coupling is small. Hence the slope $dI/dV_B$ of the Josephson branch is, for a given environment, an increasing function of the Josephson coupling. Hereafter we call supercurrent the highest value of the current on the Josephson branch. The supercurrent is 2e periodic, very small for $C_G V_G/e=0$ (modulo 2) and maximum for $C_G V_G/e=1$ (modulo 2). The change in the slope of the Josephson branch at low bias and $C_G V_G=e$ is attributed to Zener effect \cite{billangeon_prep06}. The 2e periodicity is expected from the gate dependence of the energy levels of the SCPT \cite{JoyezPhD}, which has the hamiltonian~:
\begin{eqnarray}
	H &=& \sum_n \left\{ E_C \left(n-\frac{C_G V_G}{e}\right)^2 |n\rangle \langle n| \right. \nonumber \\
	&-& \left. E_J \cos{\left(\frac{\delta}{2}\right)} \left(|n\rangle \langle n+2| + |n+2\rangle \langle n|\right) \right\}+ H_S \nonumber
\end{eqnarray}
with $E_C=e^2/2 C_\Sigma$ the charging energy ($C_\Sigma$ the total capacitance of the island), $E_J$ the Josephson energy of each junction, $\delta$ the superconducting phase difference between reservoirs and $|n\rangle$ the state with $n$ electrons on the island. $H_S$ describes a superconducting metal by the BCS theory and makes it energetically favorable to have an even number of electrons on the island, leading to an odd-even free energy difference, which value is close to $\Delta$ at low temperature \cite{Tuominen92}. The ground state of the system, if $\Delta>E_C$, consists only of paired electrons. At higher voltage $V_B$, tunneling of a Cooper pair together with QP (Josephson-QP peak) and tunneling of QP (QP step) are possible (inset of Fig. \ref{fig:figure1}). The $I(V_B,V_G)$ curve of the SCPT yields its charging energy ($E_C=65 \, \mu$eV) and the Josephson coupling of one junction ($E_J=28 \, \mu$eV).

We use a voltage biased SQUID, constituted by two Josephson junctions in parallel, named hereafter the source, as a high frequency generator to irradiate the SCPT. When the SQUID bias $V_S$ is non zero but below $2 \Delta/e$, an alternating Cooper pair current related to the AC Josephson effect runs through the SQUID. This current writes $I(t)=I_C(\Phi) \sin(\omega_J t)$ with $\omega_J/2 \pi = 2e V_S/h$ the Josephson frequency, determined by the source voltage $V_S$. The critical current $I_C(\Phi)$ is modulated by a  magnetic flux $\Phi$ through the SQUID and its maximum value $\pi \Delta/(2eR_T)$ is determined by the source's normal state resistance $R_T$ \cite{tinkham96}. The microwave photons generated by the junction are coupled to the SCPT \textit{via} the on-chip circuit. This offers a tunable frequency together with an efficient and nearly frequency independent coupling in the range 10-200 GHz, but with a very small power generated (estimated value 17.5 fW) and a not perfectly monochromatic  signal due to the emission bandwidth of the AC Josephson effect \cite{dahm69}, in particular in the dissipative environment of our setup \cite{billangeon_prep06}.

\begin{figure}
	\begin{center}
		\includegraphics[width=7.5cm]{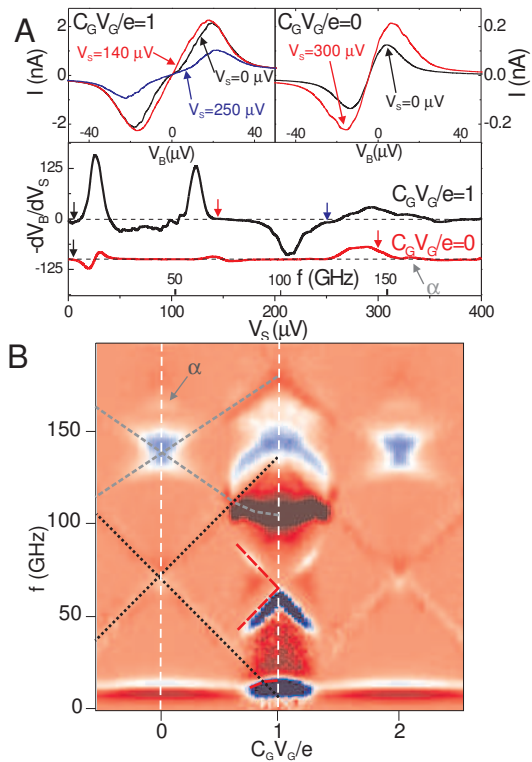}
	\end{center}
	\caption{A : Upper panels : Josephson branch of the SCPT at $C_G V_G/e=1$ (left panel) and $C_G V_G/e=0$ (right panel) and different voltage $V_S$ of the source, i.e. different Josephson frequencies. Lower panel : derivative of the voltage across the SCPT when current biased at $I=80 \, pA$ versus $V_S$ (see text) under HF irradiation for $C_G V_G/e=1$ and $C_G V_G/e=0$. The arrows correspond to voltage $V_S$ shown in the upper panels. The curves are shifted for clarity, the dashed line indicates no signal. B : Variation of $V_B$ in logarithmic scale at small current (see text) under HF irradiation and for different values of gate voltage. The two vertical dashed lines correspond to the curves shown on A, lower panel. The colored line on the left part are guides to the eye.}
	\label{fig:figure2}
\end{figure}

Below we study the effect of this irradiation on the SCPT, at maximum source power. On Fig. \ref{fig:figure2}A, upper panels, we show the modification of the Josephson branch at $C_G V_G/e=1$ and $C_G V_G/e=0$ for different source voltages $V_S$, i.e. different Josephson frequencies. For $C_G V_G/e=0$, where the supercurrent is the smallest, HF irradiation at 145 GHz ($V_S=300 \, \mu V$) leads to an increase of both the supercurrent and the slope $dI/dV_B$. For $C_G V_G/e=1$ HF irradiation leads to a global \textit{increase} of both the supercurrent and $dI/dV_B$ at 68 GHz ($V_S=140 \, \mu$V), even though the supercurrent is maximum at the chosen gate value. At higher frequency (121 GHz, $V_S=250 \, \mu$V), the Josephson branch of the SCPT is strongly reduced. The slope of the Josephson branch and the supercurrent exhibit the same qualitative dependence vs irradiation, which we relate to an induced change of the average Josephson energy of the SCPT. 
More precisely we call ''phase dependence'' the amplitude of the superconducting phase dependence of a given energy level~: $\delta E_J=\max(E(\delta))-\min(E(\delta))$. By analogy with a single Josephson junction, this quantity can be seen as the Josephson energy of the level under consideration. If due to irradiation, both the ground state of the SCPT, with a phase dependence $\delta E_{Ji}$, and an excited level, with a phase dependence $\delta E_{Jf}$, are populated with respective probability $P_i$ and $P_f$ the average phase dependence $\delta E_J^{eff}=P_i \delta E_{Ji}+ P_f \delta E_{Jf}$ will determine the slope of the Josephson branch and the supercurrent detected in the experiment.

\begin{figure}
	\begin{center}
		\includegraphics[width=7.5cm]{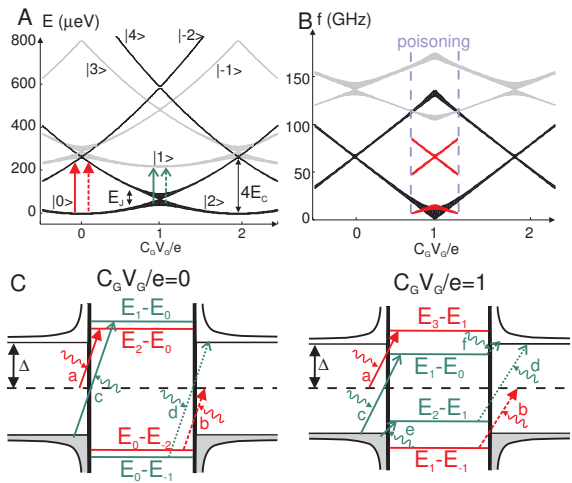}
	\end{center}
	\caption{A : Calculated spectrum of the SCPT with parameters $E_C=65 \mu V$ and $E_J=28 \mu V$. The colored arrows represent transitions induced by HF irradiation, which involve the transfer of a CP to the island (red arrow) or from the island (red dashed arrow), or the transfer of a QP to the island (green arrow) or from the island (green dashed arrow). B. Predicted gate dependence of the frequencies at which a change in the averaged phase dependence (see text) of the SCPT is expected, involving CP (black), QP (gray) tunneling or presence of an unpaired electron on the island (red). Poisoning is indicated on the figure. C~: Sketch of the QP density of states in the contacts and the energy of charge states in the SCPT at two gate values. The arrows illustrate the different photo-assisted tunneling processes possible.}
	\label{fig:figure3}
\end{figure}

In the following we focus on the slope $dI/dV_B$ of the Josephson branch, which is easier to measure, and is closely related to $\delta E_J^{eff}$ \cite{note}, since, for a given electromagnetic environment, $dI/dV_B$ is an increasing function of $\delta E_J^{eff}$. To measure $dI/dV_B$, the SCPT is current-biased with a small fixed DC current ($I=80$ pA), and the voltage $V_B$ across the SCPT is monitored. To increase sensitivity we modulate the source voltage $V_S$, and detect with a lock-in technique the modulated voltage of the SCPT, which is proportional to $dV_B/dV_S$. Figure \ref{fig:figure2}A, lower panel shows for $C_G V_G/e =1$ a strong increase of $\delta E_J^{eff}$ for frequencies around 12 GHz and 60 GHz, leading to a decrease of $V_B$ versus $V_S$. This is followed by a reduction of $\delta E_J^{eff}$ for frequencies above 100 GHz. For $C_G V_G/e =0$, HF irradiation leads to a higher $\delta E_J^{eff}$ for frequencies close to 70 GHz. Above 125 GHz $\delta E_J^{eff}$ increases. Figure \ref{fig:figure2}B shows the same data ($dV_B/dV_S$ versus Josephson frequency) as a function of $V_G$ \cite{NoteImpedance}. To interpret these results we numerically solve the hamiltonian of the system, with the parameters deduced from the transport experiment. To account for the even-odd asymmetry a free energy $\Delta$ is added to the energy of the odd charge states. This yields the energy levels of the SCPT as a function of $V_G$ and superconducting phase $\delta$. In Fig. \ref{fig:figure3}A the levels are plotted as a function of $V_G$. We include the $\delta$ dependence through a width, which represents the phase dependence $\delta E_J$ of the considered level. The absorption of a photon induces transitions between levels, followed by relaxation to the ground state, with two types of processes~: i) the absorption of one photon of energy $\hbar \omega$ allows the transfer of one Cooper pair from or to the island (transition $b$ and $a$ of Fig. \ref{fig:figure3}C at $C_G V_G/e=0$), leading to a transition from the initial energy level $E_i$ to the final energy level $E_f$, with $E_f-E_i=\hbar \omega$. The expected signal is then a peak (increase of $\delta E_J^{eff}$) or a dip (decrease of $\delta E_J^{eff}$). ii) processes which involve the transfer of one quasiparticle from the island (transition $c$ and $d$ of Fig. \ref{fig:figure3}C at $C_G V_G/e=0$). The energy balance is then $\hbar \omega \geq (E_f-E_i + \Delta)$ to allow the injection or extraction of a QP to the reservoirs (Fig. \ref{fig:figure3}B). The expected signal is then step-like, and its derivative is a peak. With these basic rules we can deduce the frequencies at which a change in the slope of the Josephson branch, due to a transition to a level with a different phase dependence $\delta E_J$, is expected. If the system is always in the lowest energy state, we get the black lines (involving transfer of CP) and gray lines (involving transfer of QP) of Fig. \ref{fig:figure3}B. This reproduces correctly the data except near $C_G V_G/e=1$, where the sharpest features are not predicted. To do so we assume that, near $C_G V_G/e=1$, the system can also be in the state with one QP on the island (state $|1\rangle$) and thus consider transitions starting from this state (transition $a,b,e$ and $f$ of Fig. \ref{fig:figure3}C at $C_G V_G/e = 1$). The calculation then reproduces accurately the frequencies and qualitatively the phase dependence of energy levels induced by the Josephson coupling. The agreement for the phase dependence is only qualitative due to intrinsic bandwidth of emission of our HF source and/or widening introduced by the photon-assisted tunneling.

We find that the calculated spectrum of the SCPT is consistent with the data up to 200 GHz. Moreover, despite the 2e periodicity versus gate of the Josephson branch, the SCPT is still affected by poisoning \cite{Aumentado04,Mannik04,Naaman06,Ferguson06}, i.e. the presence of a QP on the island for $C_G V_G \approx e$. In our case the poisoning is revealed by HF irradiation and happens for $C_G V_G/e$ between 0.7 and 1.3. Because the phase dependence of state $|1\rangle$ at $C_G V_G/e=1$ is very small, poisoning leads to a reduction of $\delta E_J^{eff}$. HF irradiation at proper frequencies allows a quicker escape of QP, leading to an increase of $\delta E_J^{eff}$. The presence of QP on the island is a common problem of SCPT, and more generally circuits based on small Josephson junctions and is a strong limitation for the coherence of these systems. We show that irradiation at sufficiently high frequencies can accelerate the escape of these QP and thus reduce the effect of poisoning. On the contrary, by using frequencies higher than $2\Delta$ we can force poisoning. In this regime the SCPT is extremely sensitive to photons and can have interesting bolometric application \cite{hergenrother94}.

We can access the dynamics of poisoning in the SCPT. To do so we assume a quantum efficiency (number of electrons transfer on the SCPT for a given number of incident photons) for photon assisted tunneling similar to the one measured in another experiment with two Josephson junctions coupled capacitively in the same environment as the present sample \cite{billangeon06}. We relate the measured change of supercurrent under irradiation to the probability for the system to be in an excited state with a QP on the SCPT, which is directly related to the time spent by the QP on the SCPT. We thus deduce a QP lifetime of 0.5 $\mu$s at $C_G V_G/e=0$ under irradiation at 145 GHz (Fig. \ref{fig:figure2}A, right upper panel) and 0.8 $\mu$s at $C_G V_G/e=1$ under irradiation at 121 GHz (Fig. \ref{fig:figure2}A, left upper panel). Those values are smaller than the ones measured for regular poisoning \cite{Naaman06,Ferguson06}. This difference is attributed to the non-equilibrium situation due to HF irradiation.

So far only transfer of one CP or one QP was considered. Actually the sequential tunneling of one QP and then one CP, followed by relaxation to the ground state is also possible. This may be the case at point $\alpha$ (Fig. 2B) where the frequency (160 GHz) is high enough to allow tunneling of one QP on the island when absorbing a photon. (transition from state $|0 \rangle$ to state $|1 \rangle$). If, before this QP leaves the island, another photon is absorbed, the SCPT can go to the charge state $|3 \rangle$, with a lower $\delta E_J$ than $|1 \rangle$, leading to a reduction of the averaged phase dependence, observed in the experiment. This effect exists only for a high enough flux of photons and indeed disappears at lower power (not shown).

In conclusion by using a Josephson junction as a tunable HF generator and by coupling it capacitively to a SCPT we perform spectroscopy on this SCPT up to 200 GHz. Up to high energy the experimental spectrum of the SCPT is consistent with the calculated energy levels with only two parameters, the charging energy and the Josephson coupling. Moreover, thanks to the high-energy of photons involved in our experiment we can reveal and reduce the poisoning of the SCPT by HF irradiation. We also demonstrate that we can induce the transfer of quasiparticle onto the SCPT, i.e. induce poisoning, and investigate the dynamics of this non-equilibrium poisoning. Our experiment shows that HF irradiation can lead not only to a decrease but also an increase of the average Josephson energy , and thus supercurrent.

\begin{acknowledgments} 
We acknowledge fruitful discussions with S. Gu\'eron. 
\end{acknowledgments}


\begin{thebibliography}{100}
\bibitem{NoteJunction}
We observe the same behavior on the Josephson junction, which has the same electromagnetic environment as the SCPT.
\bibitem{note}
Note that our simple model does not take into account the interplay of irradiation effect and Zener effect around $C_G V_G/e=1$.
\bibitem{NoteImpedance}
Note that the nearly gate independent feature seen at $V_S=20 \, \mu$V (f=10 GHz) is attributed to the change of impedance of the SCPT environment when the source is biased on the Josephson branch.
\end{thebibliography}
\end{document}